\documentclass[12pt,preprint]{aastex6}
\usepackage{psfig,epsfig,amsfonts,amsmath,amssymb,graphicx,morefloats,appendix,tensor}
\usepackage{color}
\usepackage{natbib}
\begin{document}

\slugcomment{Accepted by ApJ: November 17, 2016}

\title{ALMA Measurements of Circumstellar Material in the GQ Lup System
}

\author{Meredith A. MacGregor\altaffilmark{1}, 
David J. Wilner\altaffilmark{1},
Ian Czekala\altaffilmark{1,2},
Sean M. Andrews\altaffilmark{1},
Y. Sophia Dai\altaffilmark{3},
Gregory J. Herczeg\altaffilmark{4},
Kaitlin M. Kratter\altaffilmark{5},
Adam L. Kraus\altaffilmark{6},
Luca Ricci\altaffilmark{1},
Leonardo Testi\altaffilmark{7}
}

\altaffiltext{1}{Harvard-Smithsonian Center for Astrophysics, 60 Garden Street, Cambridge, MA 02138, USA}
\altaffiltext{2}{Kavli Institute for Particle Astrophysics and Cosmology (KIPAC), Stanford University, Stanford, CA 94305, USA}
\altaffiltext{3}{Caltech/IPAC, 1200 E. California Blvd., Pasadena, CA, 91125, USA}
\altaffiltext{4}{Kavli Institute for Astronomy and Astrophysics, Peking University, Yi He Yuan Lu 5, Haidian Qu, 100871 Beijing, China}
\altaffiltext{5}{Department of Astronomy, University of Arizona, Tucson, AZ 85721, USA}
\altaffiltext{6}{Department of Astronomy, The University of Texas at Austin, Austin, TX 78712, USA}
\altaffiltext{7}{European Southern Observatory (ESO) Headquarters, Karl-Schwarzschild-Str. 2, D-85748 Garching bei Muenchen, Germany}

\begin{abstract}

We present ALMA observations of the GQ Lup system, a young Sun-like star 
with a substellar mass companion in a wide-separation orbit.
These observations of 870~$\mu$m continuum and CO J=3--2 line emission 
with beam size $\sim0\farcs3$ ($\sim45$~AU) resolve the disk of dust and gas
surrounding the primary star, GQ Lup A, and provide deep limits on any
circumplanetary disk surrounding the companion, GQ Lup b.
The circumprimary dust disk is compact with a FWHM of $59\pm12$~AU, 
while the gas has a larger extent with a characteristic radius of $46.5\pm1.8$~AU.
By forward-modeling the velocity field of the circumprimary disk based on the 
CO emission, we constrain the mass of GQ Lup~A to be $M_* = (1.03\pm0.05)*(d/156\text{ pc})$~$M_\odot$,
where $d$ is a known distance, and determine that we view the disk at an inclination angle of $60\fdg5\pm0\fdg5$ and a 
position angle of $346\degr \pm1\degr$. The $3\sigma$ 
upper limit on the 870~$\mu$m flux density of any circumplanetary disk 
associated with GQ Lup b of $<0.15$~mJy implies an upper limit 
on the dust disk mass of $<0.04$~$M_\oplus$ for standard assumptions about optically thin emission.
We discuss proposed mechanisms for the formation of wide-separation substellar companions
given the non-detection of circumplanetary disks around GQ Lup b and other similar systems.

\end{abstract}


\section{Introduction}
\label{sec:intro}

Direct imaging surveys for extrasolar planets are revealing a surprising 
population of low-mass companions at wide-separations (semi-major axis 
$>100$~AU) \citep{cha05,luh06,laf08,ire11,krau14,bow15,krau15}.  
These substellar ($<$~40~$M_\text{Jup}$) companions present serious 
challenges to standard models of both planet and binary star formation 
\cite[e.g.][]{deb06}.
Conventional ``core accretion'' models struggle to form such massive objects 
at large semi-major axes \citep{pol96,lam12}, while core fragmentation 
and gravitational instability are difficult to arrest at low masses and 
preferentially form more massive objects \citep{bate03,krat10,jia04,bol10}.  
Another possibility is that these objects formed closer in to their host 
stars and were subsequently scattered (or migrated) outwards through dynamical 
interactions with another close in companion \citep{boss06,crida09}.  

The growing population of wide-separation companions offers a new window to
explore the processes of giant planet assembly and the subsequent formation
of moon systems.  Several of these companions exhibit line emission, as well as infrared
and ultraviolet excesses commonly associated with ongoing
accretion from ``circumplanetary'' disks \citep{seif07,sch08,bow11,bow14,bai13,zhou14}.  
There is also evidence for circumplanetary disks around planets at closer separations from photometric transit surveys \cite[J1407, see][]{mam12}.  Models of giant planet formation make testable predictions about the size,
scale height, and mass distribution of these circumplanetary disks \cite[e.g.][]{ayl09}.  Furthermore,
the properties of these disks govern the composition
and orbits of any moons that may form \citep{hel14}. 
 
One of the most prominent and best characterized examples of a system with a 
directly imaged low-mass, wide-separation companion with evidence for a circumplanetary disk is GQ Lup.  
We present new observations of 870~$\mu$m continuum 
and CO J=3--2 line emission from the GQ Lup system made with the Atacama Large 
Millimeter/submillimeter Array (ALMA).  
These new ALMA observations place a stringent upper limit on the emission from any 
circumplanetary disk surrounding GQ Lup b, and they provide strong
constraints on the geometry of the disk surrounding GQ Lup A.
We introduce the GQ Lup system in Section~\ref{sec:gqlup}. 
In Section~\ref{sec:obs}, we present the ALMA observations.
In Section~\ref{sec:results}, we describe the analysis techniques 
and results for both continuum and line emission. In Section~\ref{sec:disc},
we discuss the significance of the results on the circumprimary disk geometry, 
the limit on a circumplanetary disk, and implications for the formation 
mechanisms of wide-separation, substellar companions.

\section{The GQ Lup System}
\label{sec:gqlup}

The GQ Lup system is located in the $3\pm2$~Myr-old \citep{alc14} Lupus I cloud \citep{tac96} at a distance of $156\pm50$~pc \citep[determined from parallax, see][]{neu08}. New parallax measurements from \emph{Gaia} DR1 for stars in Lupus I, yield an average parallax of $6.4\pm0.3$~mas or $156.3\pm7.3$~pc \citep{lin16}, comparable to the earlier parallax measurements. The primary star, 
GQ Lup~A, is a classical T Tauri star \cite[spectral type K7V,][]{khar09},
with a photospheric temperature of $\sim4000-4300$~K \citep{pec13,her14,don12}.
\cite{dua08} estimate a stellar radius of $1.8\pm0.3$~$R_\odot$ and assume an effective temperature 
of $4060$~K to determine a stellar luminosity of $0.8\pm0.3L_\odot$.
Although they adopt a much higher effective temperature of $4300\pm50$~K, \cite{don12}
obtain a comparable stellar radius of $1.7\pm0.2$~$R_\odot$.
Previous estimates of the mass of GQ Lup A vary between 0.7 and 1.05~$M_\odot$, largely depending on the evolutionary models and effective temperatures used \citep{mug05,dua08,don12}.  Adopting the higher effective
temperature of $4300$~K yields a mass of $1.05\pm0.07$~$M_\odot$, the upper value in this range \citep{don12}.  Additionally, GQ Lup~A  possesses strong mid- and far-infrared excesses, indicative of a 
circumstellar disk \citep{hug94}.  \cite{dai10} marginally resolved 1.3~mm 
dust emission from the circumstellar disk using the Submillimeter
Array (SMA) and determined an outer radius of $<75$~AU. 

The substellar companion, GQ Lup b, was discovered by \cite{neu05} using the \emph{Hubble 
Space Telescope} (HST). By fitting to the broadband spectral energy distribution, \cite{zhou14}
determine that the companion has a radius of $6.5\pm2$~$R_
\text{Jup}$, an effective temperature of $2050\pm350$~K, and a luminosity of
$\text{log}L_\text{phot}/L_\odot = -2.25\pm0.24$.  The mass of this companion is uncertain, with estimates
ranging from $10-36$~$M_\text{Jup}$ \citep{mar07,seif07,neu08,lav09}. The 
projected separation of the companion from the primary star is $0\farcs7$ 
\citep{gin14}, and recent work by \cite{sch16} favors orbits with high 
eccentricity and semi-major axes $100-185$~AU.  Near-infrared spectroscopy by 
\cite{seif07} showed Pa$\beta$ line emission (equivalent width, EW~$=-3.83\pm0.12$~\AA), though subsequent observations 
by \cite{lav09} give a limit an order of magnitude lower for the same line (EW~$=-0.46\pm0.08$~\AA), 
possibly pointing to time variability of Pa$\beta$ and ongoing disk accretion. 
Optical photometery using HST shows a significant blue excess that corresponds 
to an accretion rate $\sim5\times10^{-10}$~$M_\odot$~yr$^{-1}$ \citep{zhou14}.

\section{Observations}
\label{sec:obs}

The GQ Lup system was observed with ALMA in Band 7 (870 $\mu$m) 
in a one hour (total of $\sim30$~minutes on-source) Scheduling Block (SB) 
on 2015 June 14 with 41 operational antennas and 
on 2015 June 15 with 37 operational antennas, 
using baselines that spanned 15 to 784~m.  
An additional one hour SB was executed on 2015 August 28 with 40 operational antennas,
using baselines reaching to 1574~m.  These observations are summarized in 
Table~\ref{tab:obs}, including the dates, baseline lengths, weather conditions,
and time on-source.  Overall, the weather was very good (pwv $\lesssim1.1$~mm).
The correlator was configured to optimize continuum sensitivity, while 
including both the $^{12}$CO and $^{13}$CO J $=3-2$ transitions at 345.79599 
and 330.58797~GHz, respectively.  The setup used four basebands, centered at 
331, 333, 343, and 345~GHz, in two polarizations. The basebands with the
targeted spectral lines, centered at 331 and 345~GHz, each have 3840 channels 
over a bandwidth of 1.875~GHz, while the other two basebands each have only 
128 channels over a bandwidth of 2~GHz. 
The phase center for the June observations was specified at 
$\alpha=15^\text{h}49^\text{m}12.082861$, 
$\delta=-35\degr39\arcmin05\farcs48071$ (J2000),
and the phase center for the August observations was 
$\alpha=15^\text{h}49^\text{m}12.082607$, 
$\delta=-35\degr39\arcmin05\farcs48550$.  These phase centers correspond to 
the J2000 position of the star corrected for its proper motion of 
$(-15.1,-23.4 )$~mas yr$^{-1}$.  The field of view at the center frequency 
of 338 GHz is $\sim 18\arcsec$, set by the FWHM primary beam size of the 12-m 
diameter array antennas.

The raw datasets were delivered with calibration scripts provided by ALMA staff. 
We executed these scripts for each SB using the \texttt{CASA} package 
(version 4.5.0) to generate calibrated visibilities.  Time-dependent gain variations due to atmospheric and 
instrumental effects were corrected using interleaved observations of the 
calibrator J1534-3526.  Bandpass calibration was determined from observations 
of J1517-2422.  The absolute flux calibration was derived from observations of 
Titan and Ceres, with a systematic uncertainty estimated at less than $10\%$.  
A single iteration of phase-only self-calibration was employed, after which the visibilities were
averaged into 30~second intervals.  We generated both continuum and CO line images using the multi-frequency synthesis 
\texttt{CLEAN} algorithm in \texttt{CASA}.  For spectral line imaging, the continuum level was subtracted from the
spectral windows containing emission lines.

\section{Results and Analysis}
\label{sec:results}

\subsection{Continuum Emission}
\label{subsec:continuum}

\begin{figure}[ht]
\begin{minipage}[h]{0.5\textwidth}
  \begin{center}
       \includegraphics[scale=0.75]{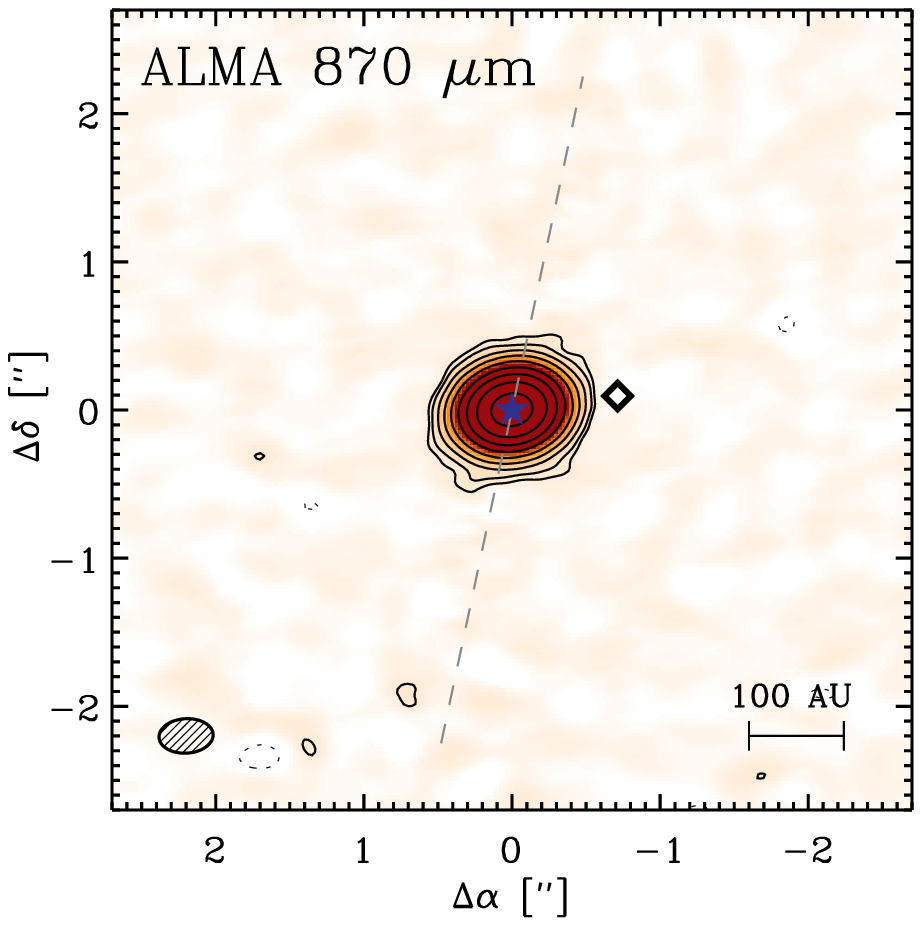}
  \end{center}
 \end{minipage}
\begin{minipage}[h]{0.5\textwidth}
  \begin{center}
       \includegraphics[scale=0.9]{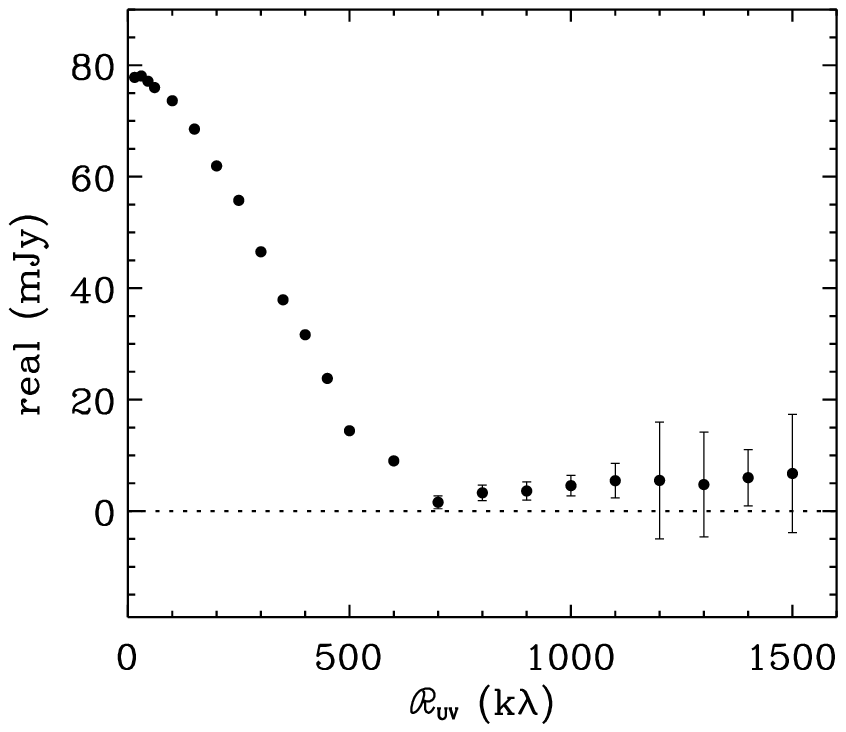}
  \end{center}
 \end{minipage}

\caption{\small  \emph{(left)} ALMA image of the 870~$\mu$m continuum emission from GQ Lup. The contour levels are in steps of [3, 6, 12, 24, 48,...]$\times50$~$\mu$Jy, the rms noise level in the image. The star symbol marks the position of the primary star, GQ Lup A, and the diamond indicates the projected location of the secondary companion, GQ Lup b.  The dashed gray line shows the position angle of the disk major axis determined by forward-modeling the CO emission and the dashed ellipse indicates the $0\farcs37\times0\farcs23$ (FWHM) synthesized beam size.
\emph{(right)} The deprojected real visibilities averaged in bins of $u, v$ distance.
}
\label{fig:cont}
\end{figure}

Figure~\ref{fig:cont} (left panel) shows the ALMA 870~$\mu$m continuum 
emission.  With robust $=0.5$ weighting, the synthesized beam size is 
$0\farcs37\times0\farcs23$ ($58\times36$~AU at 156~pc) with a position angle of 
$-87\degr$, and the rms noise level is 50~$\mu$Jy/beam.  This image reveals 
compact dust continuum emission around GQ Lup A (star symbol) and no evidence for 
emission at the position of GQ Lup b (diamond symbol).  From this non-detection, we determine 
a $3\sigma$ upper limit on the flux density of any circumplanetary disk 
surrounding the secondary of $<0.15$~mJy (assuming a point source).  
The right panel of Figure~\ref{fig:cont} shows the deprojected real 
visibilities averaged in bins of $u, v$ distance, centered on GQ Lup A
using the disk inclination and orientation determined by forward-modeling the 
CO emission (see Section~\ref{subsec:co} for a detailed description).
The profile shows a central peak and fall-off, but without the subsequent ringing that would be expected for a simple disk 
with a single radial power law surface brightness profile.
A more complicated surface brightness profile \cite[e.g. a broken power law, 
see][]{hog16} is more consistent, but a proper radiative transfer 
calculation will be needed to determine the precise radial profiles of the 
disk surface density and temperature.  The deprojected imaginary visibilities
are consistent with zero, as is expected for a symmetric structure.

By fitting a simple two dimensional Gaussian to the continuum image, we obtain 
a total flux density for the circumprimary disk of $77.8\pm0.2$~mJy, 
consistent with previous interferometric and single dish millimeter flux 
measurements.  \cite{dai10} measure a flux density of $25\pm3$~mJy at 1.3~mm 
with the SMA.  \cite{nuer97} measure a flux density of $38\pm7$~mJy at 1.25~mm 
with the SEST bolometer.  If we extrapolate our ALMA measurement using a typical 
spectral index for T Tauri stars of $2.4\pm0.5$ \citep{and13}, we obtain a flux density at 1.3~mm of $29.7\pm5.4$~mJy, 
in good agreement with both previous flux density measurements within their 
uncertainties.  Given this flux density, the GQ Lup circumstellar disk is brighter 
than $\sim70\%$ of other Lupus sources with spectral types K4$-$M1 \citep{ans16}.  The major axis FWHM of the continuum emission (deconvolved 
from the beam) is $0\farcs38\pm0\farcs07$.  At a distance of 156~pc, this 
gives a characteristic size for the primary disk of $59\pm12$~AU, again 
comparable to the results of \cite{dai10}, who derived an outer radius for the disk of 
$25-50$~AU ($\sim50-100$~AU in diameter).  Longer baseline observations with higher angular resolution are 
needed to better constrain the location and sharpness of the dust disk edges, 
and to probe for any substructure that might betray the presence of an 
additional inner companion in the system.

For optically thin emission, we can make a simple estimate of the total dust mass 
($M_\text{dust}$) for the circumprimary disk given the observed total 
flux density \citep{hil83}:

\begin{equation}
M_\text{dust} = \frac{F_\nu D^2}{\kappa_\nu B_\nu(T_\text{dust})}.
\label{eq:mdust}
\end{equation}

\noindent Here, $B_\nu(T_\text{dust})$ is the Planck function at the dust temperature, $T_\text{dust}$, and $\kappa_\nu$ is the dust opacity.  For consistency with \cite{bow15} and \cite{ans16}, we adopt the frequency-dependent dust opacity $\kappa_\nu = 10(\nu/10^{12}\text{ Hz})$~cm$^{2}$~g$^{-1}$ from \cite{beck90}. At 870~$\mu$m, the dust opacity is 
$\kappa_\nu=3.4$ cm$^{2}$ g$^{-1}$.  To estimate the dust temperature, we use 
the dust temperature-stellar luminosity relationship of \cite{and13}: 
$T_\text{dust}=25(L/L_\odot)^{1/4}$.  For GQ Lup A, this relation yields 
$T_\text{dust} =24\pm8$~K.  The resulting dust mass is 
$15.10\pm0.04$~$M_\oplus$.

Similarly, we can use the $3\sigma$ upper limit on the flux density of a 
circumplanetary disk around the companion GQ Lup b to place an upper limit on 
the potential dust mass.  Given its low luminosity, we assume that 
heating of a circumplanetary disk around GQ Lup~b is dominated by the primary 
star, GQ Lup~A, rather than by the companion itself.  
If we assume that the orbit of GQ Lup b and the circumprimary disk are coplanar, the radiative equilibrium temperature at the position of GQ Lup b ($\sim220$~AU, see Section~\ref{subsec:geom} for discussion),
is $18\pm2$~K.  Taking this value as a representative dust temperature for 
our analysis, the resulting $3\sigma$ upper limit on the dust mass is 
$M_\text{dust}<0.04$~$M_\oplus$.  For a gas-to-dust ratio of 100, this implies 
a total circumplanetary disk mass of $M_\text{tot}<4$~$M_\oplus$ or 
$<0.04-0.13\%$ the mass of the companion itself (for companion masses of 
$10-36$~$M_\text{Jup}$).  This estimate of the disk dust mass is 
sensitive to both the assumed dust opacity, $\kappa_\nu$, and the characteristic dust temperature, $T_\text{dust}$.  
\cite{vandP16} derive a temperature-luminosity relationship for spectral types 
M5 and later (assuming different prescriptions for disk flaring and opacity than \cite{and13}): $T_\text{dust}=22(L/L_\odot)^{0.16}$.  Given a luminosity of 
$\sim0.006$~$L_\odot$ for GQ Lup b, this relationship implies a dust 
temperature of $\sim10$~K.  If we take $T_\text{dust}=10$~K, instead, then 
$M_\text{dust}<0.14$~$M_\oplus$ and $M_\text{tot}<14$~$M_\oplus$.  
Even for this low temperature the total disk mass is $\mathbf{\lesssim0.1-0.3\%}$ of the
companion mass.  Any possible viscous heating of the disk \cite[e.g.][]{ise14} is neglected
here because of the low measured accretion rate, $\sim5\times10^{-10}$~$M_\odot$~yr$^{-1}\sim0.5$~$M_\text{Jup}$~Myr$^{-1}$.  If we take the measured accretion rate 
together with a disk mass of $\lesssim4-14$~$M_\oplus$ for the circumplanetary 
disk, this yields an expected lifetime of $\sim2\times10^4-1\times10^5$~years, 
shorter than the age of the GQ Lup system of $\sim2$~Myr.
The total dust mass of the circumplanetary disk also places constraints on the possibility
of moon formation around the companion.  In our own Solar System, the total mass of the moons of Jupiter,
Saturn, and Uranus are all $\sim10^{-4}$ the mass of their host planet \citep{can06}.  The dust content
of the GQ Lup b disk is at least a factor of six lower than this moon-planet mass ratio, making it difficult to form gas giant moons.  However, in the model of \cite{can06} satellites form in a circumplanetary disk during the final stages of growth of the host planet, so we cannot rule out the future formation of rocky moons.

If we assume optically thick dust emission for the circumplanetary disk,
then we can derive an upper limit on its size.
In this limit, the intensity, $I_\nu$, 
is approximately $B_\nu(T_\text{dust})$.  Thus,

\begin{equation}
R_\text{dust} = \sqrt{\frac{F_\nu D^2}{\pi B_\nu(T_\text{dust})}}.
\end{equation}

\noindent Given the upper limit of $F_\nu < 0.15$~mJy and a dust temperature 
of 18~K, $R_\text{dust}$ must be $<1.1$~AU.  For comparison, the Hill radius of GQ Lup b assuming
a semi-major axis of $\sim100$~AU and an eccentricity of $\sim0.2$ is $R_\text{Hill}\sim12-19$~AU (for companion masses between $10-40$~$M_\text{Jup}$).  Thus, this small disk size may be compatible
with numerical simulations of circumplanetary accretion disks that are thick, dense, 
and truncated at a few tenths of the Hill radius ($R\sim0.3-0.4R_\text{Hill}$) by the gravity of the central star
\cite[e.g.][]{ayl09,mar11,szu16}.

\subsection{$^{12}$CO and $^{13}$CO Emission}
\label{subsec:co}

\begin{figure}[ht]
\begin{minipage}[h]{0.5\textwidth}
  \begin{center}
       \includegraphics[scale=0.75]{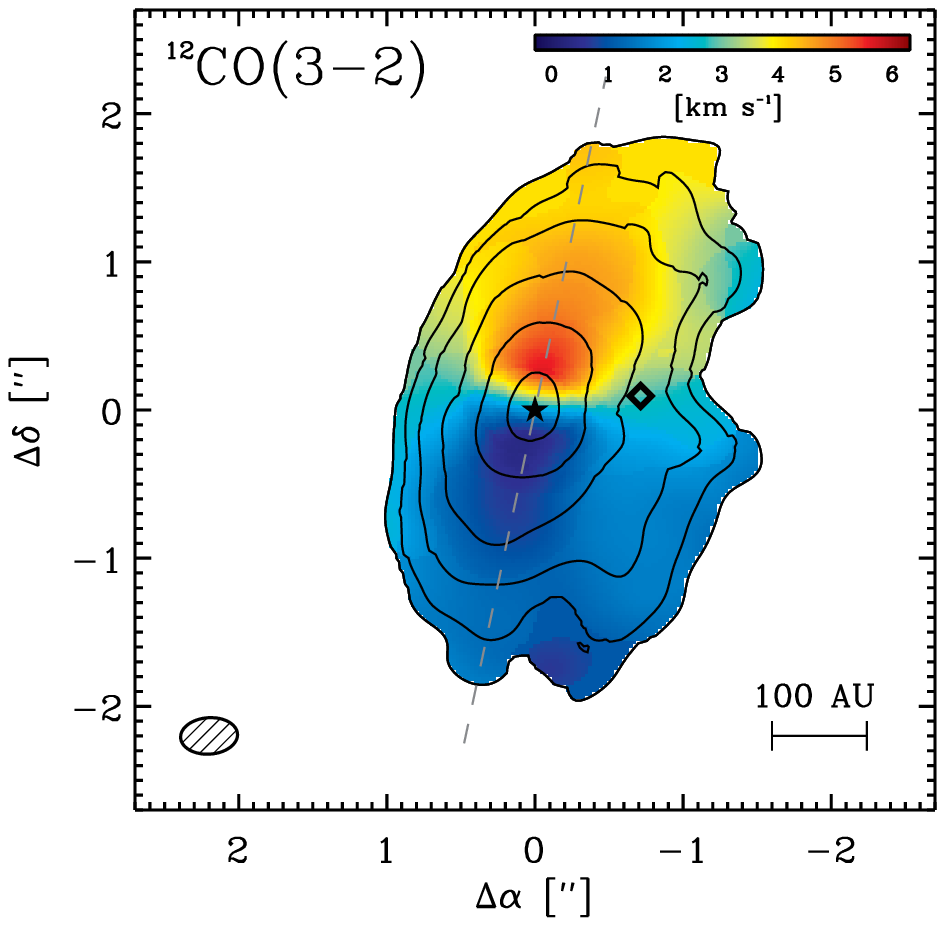}
  \end{center}
 \end{minipage}
\begin{minipage}[h]{0.5\textwidth}
  \begin{center}
       \includegraphics[scale=0.75]{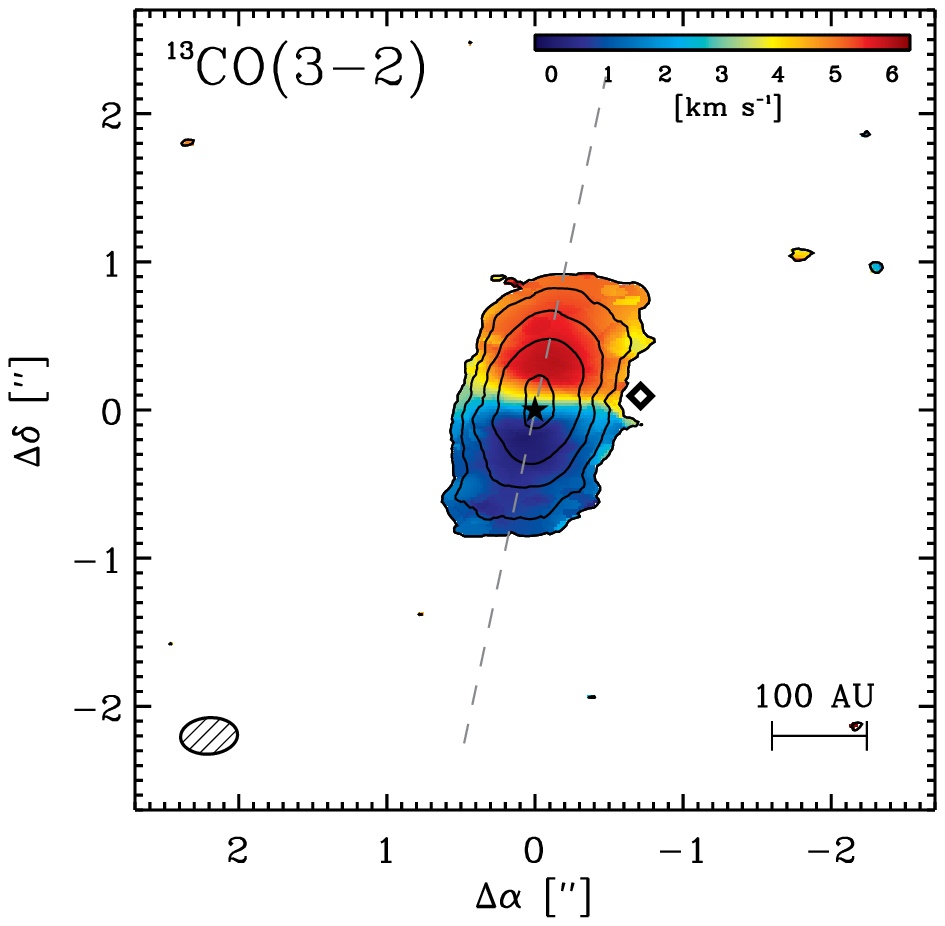}
  \end{center}
 \end{minipage}
\caption{\small  \emph{(left)} The $^{12}$CO $J = 3-2$ moment maps for the GQ Lup A circumstellar disk. The zeroth moment (velocity-integrated intensity) map is indicated by contours in steps of [3, 6, 12, 24, 48,...]$\times11$~mJy~km~s$^{-1}$~beam$^{-1}$, the rms noise level in the image.  The first moment (intensity-weighted velocity) is shown in color with a scale bar for reference.
\emph{(right)}  The $^{13}$CO $J = 3-2$ moment maps for the GQ Lup disk. The zeroth moment map is overlaid with contours in steps of [3, 6, 12, 24, 48,...]$\times6.5$~mJy~km~s$^{-1}$~beam$^{-1}$, the rms noise level in the image.  Again, the first moment is shown in color with a scale bar for reference.
In both panels, the star symbol marks the position of the primary star and the diamond indicates the projected location of the secondary companion.  The dashed gray line shows the position angle, $PA=346\degr$, of the disk major axis determined from modeling and the dashed ellipse indicates the $0\farcs37\times0\farcs23$ (FWHM) synthesized beam size.
}
\label{fig:co}
\end{figure}

\begin{figure}[ht]
\begin{minipage}[h]{1.\textwidth}
  \begin{center}
       \includegraphics[scale=0.8]{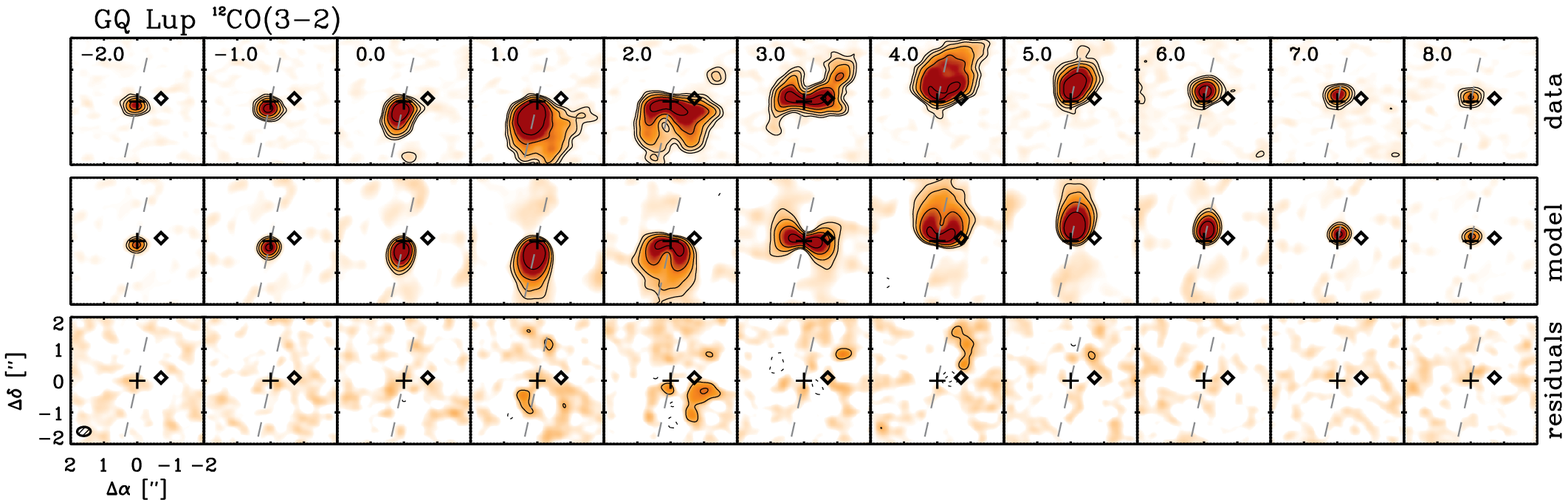}
  \end{center}
 \end{minipage}
\begin{minipage}[h]{1.\textwidth}
  \begin{center}
       \includegraphics[scale=0.8]{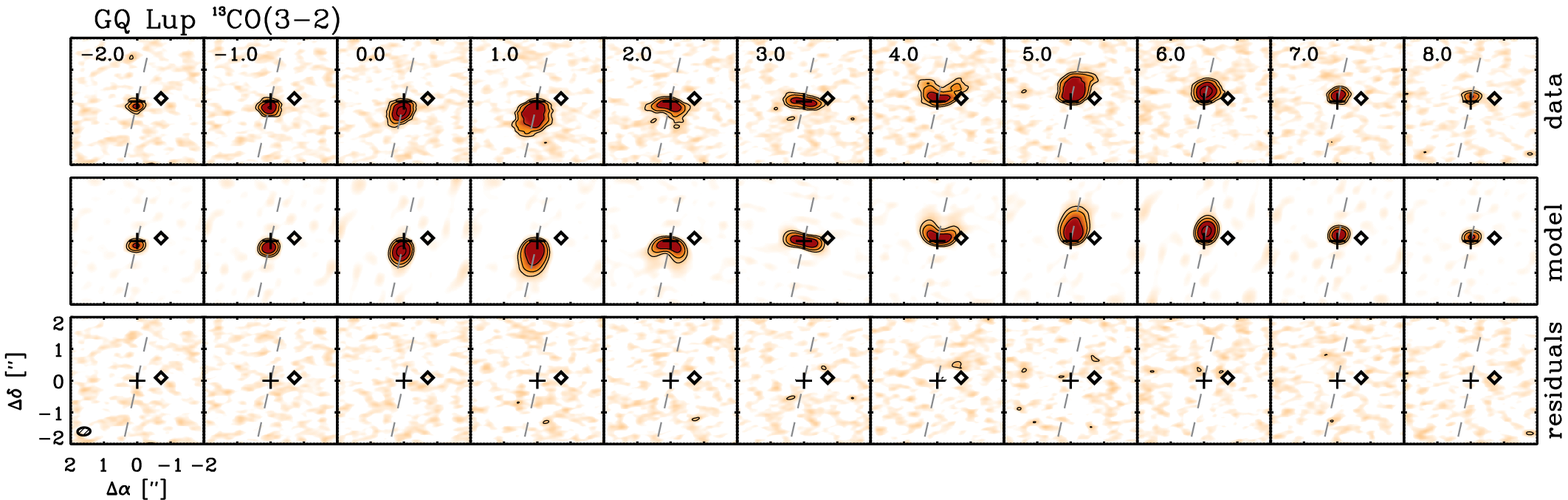}
  \end{center}
 \end{minipage}
\caption{\small  Channel maps (top row), the best-fit model (model row), and the imaged residuals (bottom row) for both the $^{12}$CO (top) and $^{13}$CO (bottom) $J=3-2$ emission for GQ Lup.  Contours for all panels are in steps of $[3, 6, 12, 24, 48,...]\times$ the rms noise level in the image, with an rms of 11~mJy/beam and 6.5~mJy/beam for the $^{12}$CO and $^{13}$CO images, respectively.  The ellipse in the lower left corner of both bottom leftmost panels indicates the $0\farcs37\times0\farcs23$ (FWHM) synthesized beam size.  Each channel is 1~km/s wide with the LSR velocities labeled in the upper left corner of each panel.
}
\label{fig:chan}
\end{figure}

Figure~\ref{fig:co} shows the velocity-integrated intensity ($0^\text{th}$ 
moment) overlaid as contours on the intensity-weighted velocity ($1^\text{st}$ 
moment) for both the $^{12}$CO and $^{13}$CO emission (left and right panels,
respectively).  Both maps show a clear pattern of Keplerian rotation, seen more
explicitly in the channel maps shown in Figure~\ref{fig:chan} (top: $^{12}$CO, 
bottom: $^{13}$CO).  Only the central 11 channels are shown for each line, where
emission is clearly resolved at $>3\sigma$.  For the $^{12}$CO image, the typical rms in a given channel
is 11~mJy/beam higher due to calibration issues in the spectral window containing $^{12}$CO for two of the scheduling blocks that were dealt with by ALMA staff).  The integrated and peak intensity are 14.5~Jy~km~s$^{-1}$ and 
1.43~Jy/beam ($130\sigma$), respectively.  For the $^{13}$CO image, the typical rms 
noise is 6.5~mJy/beam.  The integrated and peak intensity are 1.76~Jy~km~s$^{-1}$ and 
0.35~Jy/beam ($54\sigma$), respectively.  The systemic velocity in the LSRK 
frame is $3.00\pm0.01$~km~s$^{-1}$, and corresponds to 
$-2.88\pm0.01$~km~s$^{-1}$ in the barycentric frame.  \cite{sch16} recently 
derived a comparable systemic velocity for the primary of 
$-2.8\pm0.2$~km~s$^{-1}$ from near infrared observations using the CRIRES 
instrument on the VLT.

The CO emission morphology does not show any indication of truncation of the 
circumprimary gas disk due to the companion, GQ Lup b. Both the $^{12}$CO and 
$^{13}$CO emission appear largely symmetric in their spatial distribution
across the disk major axis (position angle $\sim348\degr$).  There is an
indentation and compact $>6\sigma$ emission peak visible northwest of the star in the $^{12}$CO 
moment and channel maps with velocities between 2 and 3~km~s$^{-1}$. Extended interstellar
molecular cloud material was seen by \cite{vanK07} in single dish $^{12}$CO 
emission towards GQ Lup with $v_\text{LSRK} \sim 4-5$~km~s$^{-1}$.  Although the 
velocities of the observed structure and the extended interstellar component do not match exactly, 
it is plausible that the $^{12}$CO ALMA images of the circumprimary disk 
are affected by contamination from ambient cloud emission.

In order to determine a dynamical mass for GQ Lup A and to characterize the 
gas disk geometrical properties, 
we forward-model the ${}^{12}$CO and ${}^{13}$CO molecular line emission using 
the \texttt{DiskJockey} package\footnote{Open source and freely available at 
\url{https://github.com/iancze/DiskJockey} under an MIT license.} 
\citep{cze15}. We adopt a simple parametric model of disk structure, which uses a self-similar
surface density profile \citep{lyn74} described by a characteristic radius, $r_c$, and total gas mass, $M_\text{gas}$:

\begin{equation}
\Sigma = \Sigma_c \left(\frac{r}{r_c}\right)^{-1}\text{exp}\left[-\left(\frac{r}{r_c}\right)^2\right].
\end{equation}
 
\noindent Here, $\Sigma_c$ is a normalization given by $e\times\Sigma(r_c)$, $M_\text{gas} = X_\text{CO}\Sigma_c(2\pi r_c^2)$, and $X_\text{CO}$ is the fractional abundance of CO (assumed to be constant throughout the disk). The disk is assumed to be vertically isothermal and in hydrostatic equilibrium,
with a radial power law index, $q$, and a normalization at 10~AU, $T_{10}$:

\begin{equation}
T = T_{10}\left(\frac{r}{10\text{ AU}}\right)^{-q}.
\end{equation}

\noindent The velocity field is assumed to be Keplerian with systemic velocity, $v_\text{sys}$, and is dominated by the 
stellar mass, $M_\ast$. Non-thermal (turbulent) line broadening is denoted by a constant velocity width, $\xi$.
We also include two offsets in both RA and DEC, $\Delta\alpha$ and $\Delta\delta$, respectively.
The posterior probability of the model parameters is 
evaluated in the following manner: (1) sky-images of a given disk structure are
generated using the \texttt{RADMC-3D} radiative transfer program \citep{dul12},
Fourier transformed, and sampled at the $u, v$ locations corresponding to the 
ALMA baselines, and (2) the model visibilities are then evaluated using a 
$\chi^2$ likelihood function which incorporates the statistical weights on each 
visibility measurement. This generative model allows us to fully explore the 
uncertainties in each parameter as well as determine the one dimensional 
marginalized probability distribution on stellar mass. Although more 
sophisticated models of disk structure are desirable, this simple model has 
been proven to yield accurate stellar masses, as confirmed by comparison with 
measurements of circumbinary disks around spectroscopic binaries 
\citep{ros12,cze15,cze16}. Further details of the modeling framework can be 
found in \cite{cze15}.

The best-fit parameter values and their $68\%$ uncertainties are listed in 
Table~\ref{tab:co}.  Figure~\ref{fig:chan} shows the channel maps for the 
data (top), best-fit model imaged like the data (middle), and resulting 
residuals (bottom) for both the $^{12}$CO (top grouping) and $^{13}$CO (bottom 
grouping) emission.  The results for both lines are consistent, although the 
$^{12}$CO fits may be biased by the cloud contamination evident in the 
residuals (see channels with velocities between $1-4$~km~s$^{-1}$).  As a 
result, we focus on the best-fit parameters from the $^{13}$CO modeling.

The models imply a stellar mass, $M_*$, for GQ Lup A of 
$1.03\pm0.15$~$M_\odot$. The quoted uncertainty on the mass includes the 
significant uncertainty in the distance ($\pm50$~pc) added in quadrature.  
At a known distance, $d$, the constraint on the stellar mass can be recast
as $M_* = (1.03\pm0.05)*(d/156\text{ pc})$, where the formal uncertainty on $M_*$ is $\sim5\%$ including
systematic uncertainties estimated from more complex models \cite[e.g. vertical structure, see][]{ros13}.  
Previous estimates of the stellar mass of GQ Lup A from the 
literature are mostly lower than our determination, ranging between 0.7 and 1.05~$M_\odot$ \cite[see discussion in Section~\ref{sec:gqlup},][]{mug05,dua08,don12}.  The discrepancy in mass estimates results
largely from differences in stellar evolutionary models and uncertainty in the effective temperature.
Given this result and previous work \citep{cze15,cze16,ros13}, ALMA can play a substantial role in precisely measuring the masses of large samples of young stars, providing constraints on evolutionary models.  Much work has been done to determine allowable orbits for the companion, GQ Lup b, all of which assume a stellar mass of 0.7~$M_\odot$ 
\citep{gin14,pea15,sch16}.  \cite{pea15} define a criteria for a bound orbit, 
$B<1$, where $B\propto(M/M_\odot)^{-1}$.  This new determination of the 
stellar mass of GQ Lup A may prove relevant for constraining allowable orbits 
of the secondary.

The characteristic radius and total gas mass for the best-fit model to the $^{13}$CO emission are 
$46.5\pm1.8$~AU and log$M_\text{gas}/M_\odot = -3.67\pm0.05$, respectively.  
We also compare our $^{12}$CO and $^{13}$CO integrated line intensities to the model
grids of \cite{wil14}, which predict a gas mass between $10^{-4} - 10^{-3}$~$M_\odot$, consistent with
our modeling results.  By combining this gas mass with the 
total dust mass determination from Section~\ref{subsec:continuum}, we can 
calculate the gas-to-dust ratio for the circumstellar disk around GQ Lup A 
to be $4.7\pm0.5$.  This result is well below ISM gas-to-dust ratios, but 
is comparable to measurements made by \cite{ans16} for circumstellar disks 
around other T Tauri stars in Lupus with similar stellar masses.  
In fact, nearly all of the detected Lupus disks are inferred to have 
gas-to-dust ratios well below 100.  A significant caveat to our derived gas mass
is that it depends inversely on the CO/H$_2$ abundance ratio, which 
we assume to be ISM-like $\sim10^{-4}$. Furthermore, recent work by 
\cite{mio16} suggests that a more complex analysis is required to 
accurately determine disk gas masses.

\section{Discussion}
\label{sec:disc}

We have performed interferometric observations of the GQ Lup system at 870~$\mu$m with ALMA and detected both continuum and $^{12}$CO and $^{13}$CO $J=3-2$ line emission.  The continuum image reveals compact dust emission surrounding the primary star, but no emission at the position of the secondary companion.  We place a robust $3\sigma$ upper limit on the flux of a circumplanetary disk surrounding the companion of $<0.15$~mJy.  We use the Keplerian velocity field as determined by the line emission data to estimate the mass of the primary star, $M_*=1.03\pm0.15$~$M_\odot$, and the geometry of the circumprimary disk.  We now use this new information to discuss the geometry of the circumprimary disk and implications for formation scenarios of massive companions on wide orbits.

\subsection{Circumprimary Disk Geometry}
\label{subsec:geom}

The CO emission models place tight constraints on the geometry of the 
circumprimary disk through the inclination and position angle.  For the 
$^{13}$CO emission, the best-fit inclination is $i = 60\fdg5\pm0\fdg5$ 
and position angle is $PA=346\degr \pm1\degr$.  There are discrepancies in 
the literature over the inclination angles of the stellar rotation axis and 
circumprimary disk for the GQ Lup system.  \cite{bro07} combine photometric 
rotation period monitoring with a previous measurement of $v\text{sin}i$ from 
HARPS \citep{gue05} to determine the inclination of the star's 
rotational axis to be $i=27\degr\pm5\degr$, much lower than the inclination
of the circumstellar disk. In contrast, \cite{dua08} derive a higher 
inclination of $53\degr\pm18\degr$ from spectrophotometric data taken with 
the 1.52 m ESO telescope in La Silla.  Using high resolution VLT/CRIRES spectra 
of CO emission from GQ Lup, \cite{pon11} find a best-fit disk inclination of $65\degr\pm10\degr$.
Our analysis agrees with these later determinations of the disk inclination and suggests that the 
disk inclination is significantly higher than previously estimated for the star.

Assuming that the orbit of GQ Lup b is also coplanar 
with the circumprimary disk implies that the current physical separation of GQ Lup b could 
be as large as $\sim220$~AU.  \cite{sch16} and \cite{gin14} (assuming a 
stellar mass of 0.7~$M_\odot$) propose three families of orbital solutions for 
GQ Lup b: 1) semi-major axis $\sim100$~AU, $i\sim57\degr$, eccentricity $\sim0.15$, 
2) semi-major axis $<185$~AU, $28\degr<i<63\degr$, eccentricity 0.2 to 0.75, 
and 3) semi-major axis $>300$~AU, $52\degr<i<63\degr$, eccentricity $>0.8$.  
More specifically, they note that orbits with lower eccentricities between $0.1-0.4$ have high inclinations
between $48\degr-63\degr$.  Given the apparent discrepancy between these high inclinations and the assumed low inclination of the circumstellar disk ($i\sim27\degr$), \cite{sch16} and \cite{gin14} suggest that GQ Lup b was likely scattered to its current position since {\em in situ} formation would result in a low eccentricity orbit near the plane of the circumstellar disk.  
Our new robust measurement of the circumstellar disk inclination relieves
some of this tension and does not exclude an {\em in situ} formation, since an inclination of $60\fdg5\pm0\fdg5$ is well within
the range determined for low eccentricity orbits.

\subsection{Comparison to other Young Substellar Objects}
\label{subsec:planet}

\begin{figure}[ht]
\centerline{\psfig{file=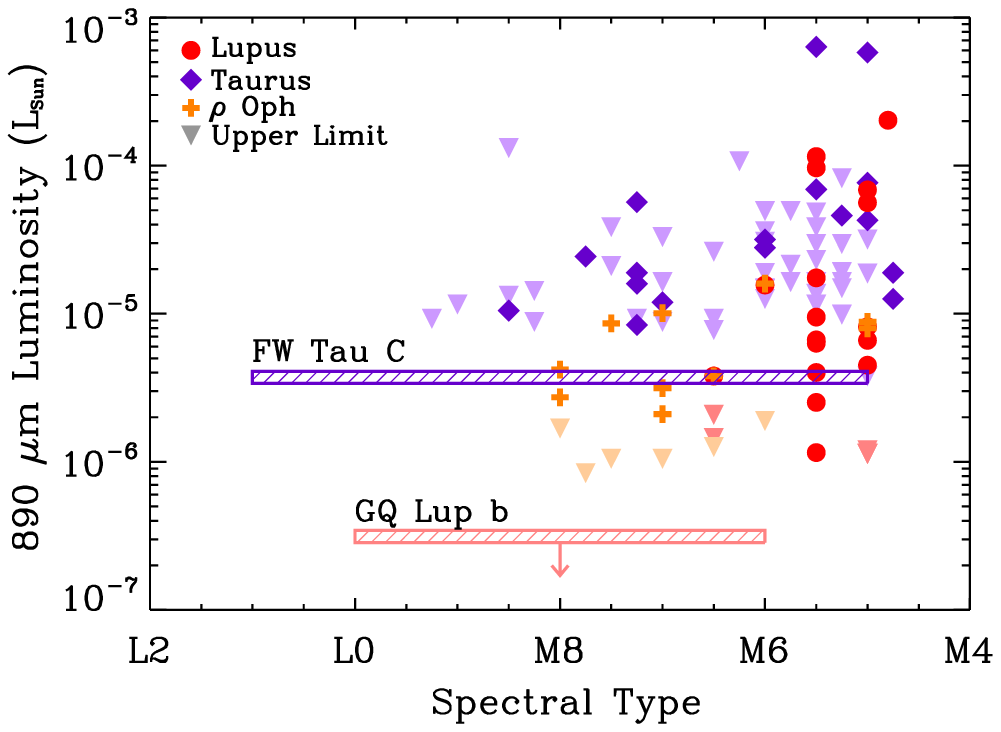,width=10cm,angle=0}}
\caption{\small (Sub)millimeter dust luminosities as a function of spectral type for the young ($\sim2$~Myr-old) Lupus \cite[red circles,][]{ans16}, Taurus \cite[purple diamonds,][]{and13,ric14}, and $\rho$ Ophiucus \cite[orange crosses,][]{tes16} star forming regions.  The upside down triangles indicate $3\sigma$ upper limits.  Our upper limit for GQ Lup b is nearly an order of magnitude lower than the previous ALMA measurement of a circumplanetary disk surrounding FW Tau C \citep{krau15}.
}
\label{fig:compare}
\end{figure}

The $3\sigma$ dust mass upper limit we obtain for GQ Lup~b is lower than 
previous circumplanetary disk mass constraints obtained with ALMA. 
\cite{bow15} observed GSC 6214-210, a $5-10$~Myr-old system with a 
$\sim15$~$M_\text{Jup}$ companion at a separation of $\sim320$~AU and did not detect dust emission surrounding 
either the primary or secondary; they place an upper limit on the 
circumplanetary dust mass of $<0.15$~$M_\oplus$ or $<0.3\%$ of the companion 
mass.  However, a non-detection of millimeter dust emission around both the 
primary and the secondary is consistent with the results of a large survey of 
the TW Hya association, which found dust masses for similar late spectral 
type objects of $\lesssim10^{-2}$~$M_\oplus$ \citep{rod15}.  In older 
systems like these, it is possible that the effects of grain growth and drift 
have depleted the disks of grains that are emissive at millimeter wavelengths.
In contrast, ALMA observations of the younger, 2~Myr-old FW Tau system 
\citep{krau15} detected significant dust emission surrounding the 
$<40$~$M_\text{Jup}$ companion at $\sim330$~AU \citep{cac15}, implying a circumplanetary dust mass of 
$1-2$~$M_\oplus$.  However, the spectral energy distribution, especially at near-infrared
wavelengths, suggests that FW Tau C is degenerate between a planetary mass object and a very low mass 
star or brown dwarf (spectral type M5$-$M8) with an edge-on disk \citep{bow14}.  
Figure~\ref{fig:compare} compares our ALMA constraint on the dust luminosity 
of a circumplanetary disk around GQ Lup b to the previous constraints on 
FW Tau C from \cite{krau15}.  Also included in Figure~\ref{fig:compare} are 
previous (sub)millimeter measurements for sources with spectral types M5 and 
later from surveys of the young ($\sim2$~Myr-old) Lupus, Taurus, and 
$\rho$ Ophiucus star-forming regions \citep{ans16,and13,ric14,tes16}.  All of the 
dust luminosities were calculated from measurements of the 890~$\mu$m flux 
density. To construct this plot, we assumed the mean distance for each star 
forming region to be the following: $156\pm50$~pc \cite[Lupus I, II, IV,][]{neu08}, 
$200\pm50$~pc \cite[Lupus III][]{com08},
$140\pm20$~pc \cite[Taurus,][]{tor12}, 
and $135\pm8$~pc \cite[$\rho$ Ophiucus,][]{mam08}. The ALMA limit for GQ Lup b 
is nearly an order of magnitude lower than the detections from these other 
large surveys.  This wide spread in dust luminosity for similar spectral type 
objects shows that there is a wide range of evolutionary outcomes for 
circumstellar disks at these young ages.

\subsection{Implications for Formation Scenarios}

The growing sample of systems with deep millimeter observations and 
corresponding limits on circumplanetary disk masses allows us to speculate 
on proposed formation mechanisms of such systems \citep{deb06,dai10}.
One possibility is that these wide-separation substellar companions formed 
{\em in situ} through core fragmentation or gravitational instability.  
However, models predict that companions formed through these mechanisms 
should be surrounded by massive circumplanetary disks that persist over 
several Myrs by accreting material from the disk of the parent star 
\citep{stam15,vor10,bol09}.  Another possibility is that these substellar 
companions formed much closer in to the primary star and were later scattered 
outward through dynamical interactions with another massive body 
\citep{boss06,crida09,sch09}.  Such chaotic events are likely to disrupt or 
destroy any circumplanetary disk surrounding the companion, since the closest 
approach is $<<R_\text{Hill}$. However, a recent survey by \cite{bry16} of eight wide separation planetary mass companions ruled out the presence of $<7$~$M_\text{Jup}$ inner companions in these systems
at separations of $15-50$~AU, suggesting that scattering may not be a dominant mechanism
for the formation of wide separation companions.  It is also possible that such systems formed
through the standard binary fragmentation route \citep{fish04,off10,bate12}, where turbulent
fragmentation and orbit evolution can result in wide-separation, unequal mass binary systems.

While the null detection of a circumplanetary disk around GQ Lup b ($M_\text{dust} <0.04$~$M_\oplus$) argues against {\em in situ} formation, its orbital parameters are still consistent with such a model. Indeed, the models of \cite{gin14} and \cite{sch16} do not exclude low eccentricity orbits as would be expected for a planet-like formation within a larger protoplanetary disk. The morphology of the GQ Lup A disk points against a scattering origin for the companion. There is no observational evidence for a sharp inner edge or cavity indicative of an additional companion in the system that may have scattered GQ Lup~b out to 
its current position. Observations with higher angular resolution are needed 
to probe for any substructure in the circumprimary disk that would signify 
the influence of an additional companion, or features that may have resulted 
from a previous scattering event.

\cite{sch16} measure the projected rotational velocity of GQ Lup b to be 
$5.3^{+0.9}_{-1.0}$~km~s$^{-1}$, making it a slow rotator compared to the 
giant planets in the Solar System and the recent measurement of 
$\beta$ Pictoris b \citep{snel14}.  Objects formed through gravitational 
instability or core fragmentation seem to follow a spin-mass trend, where 
higher mass objects rotate faster than lower mass objects.  The unusually
slow spin of GQ Lup b could point to a different formation scenario, but, 
as \cite{sch16} point out, GQ Lup b is still quite young and will likely 
spin up over time, making its slow spin less discrepant.

\section{Conclusions}
\label{sec:conclusions}

We present new ALMA observations of 870 $\mu$m dust continuum and CO J=3-2 line 
emission from the GQ Lup system.  These observations resolve the circumstellar 
disk surrounding GQ Lup A, and provide a deep upper limit on any emission 
from a circumplanetary disk surrounding GQ Lup b. The main results are as 
follows.

\begin{enumerate}

\item The circumprimary disk appears compact with a FWHM of 
$59\pm12$~AU.  Given the total flux density and assuming optically 
thin emission, we determine a total dust mass of 
$M_\text{dust} = 15.10\pm0.04$~$M_\oplus$.  

\item There is no indication that the circumprimary disk traced by $^{12}$CO 
and $^{13}$CO emission is truncated or affected by the presence of the 
companion, GQ Lup b.  The characteristic radius of the $^{13}$CO emission
is $46.5\pm1.8$~AU, more extended than the dust disk.
By forward-modeling the Keplerian velocity field,
we robustly constrain both the mass of the primary star, 
$M_* = (1.03\pm0.05)*(d/156\text{ pc})$, and the geometry of the circumprimary disk, 
$i = 60\fdg5\pm0\fdg5$ and $PA=346\degr \pm1\degr$. An inclination of 
$i=60\degr$ is significantly higher than previous estimates of $20-30\degr$.  
If the companion orbit is coplanar with the circumprimary disk, then 
this high inclination implies that the current physical separation of the secondary is 
$\sim220$~AU. 

\item We determine a robust $3\sigma$ upper limit on the flux density of any 
circumplanetary disk surrounding GQ Lup b of $<0.15$~mJy.  If we assume 
optically thin emission, then this implies an upper limit on the dust mass of 
$M_\text{dust}<0.04$~$M_\oplus$.  This limit is an order of magnitude lower 
than previous ALMA measurements for circumstellar disks around M5 and later 
sources of similar ages ($\sim2$~Myr).  In the optically thick limit, we can 
instead derive an upper limit on the radius of the circumplanetary disk of 
$R_\text{dust}<1.1$~AU.  

\item Since models of {\em in situ} formation of wide-separation, substellar 
companions through core fragmentation or gravitational instability predict 
massive circumplanetary disks that persist for several Myrs, the lack of 
detections of such massive disks disfavors these formation scenarios.

\end{enumerate}

Millimeter observations of additional systems with young substellar companions 
are needed to characterize the disk properties and to assess whether or not the
features of the GQ Lup system are typical of the whole population.  In addition, higher angular 
resolution is needed to probe for any substructure in circumprimary disks, like GQ Lup, 
that could indicate the presence of additional companions involved in dynamical evolution.

\acknowledgements
M.A.M acknowledges 
support from a National Science Foundation Graduate Research Fellowship 
(DGE1144152) and from NRAO Student Observing Support.  
This paper makes use of the following ALMA data: ADS/JAO.ALMA \#2013.1.00374.S. 
ALMA is a partnership of ESO (representing its member states), NSF (USA) and 
NINS (Japan), together with NRC (Canada) and NSC and ASIAA (Taiwan) and KASI 
(Republic of Korea), in cooperation with the Republic of Chile. The Joint ALMA 
Observatory is operated by ESO, AUI/NRAO and NAOJ. The National Radio Astronomy
Observatory is a facility of the National Science Foundation operated under 
cooperative agreement by Associated Universities, Inc. 
This work has also made use of data from the European Space Agency (ESA)
mission {\it Gaia} (\url{http://www.cosmos.esa.int/gaia}), processed by
the {\it Gaia} Data Processing and Analysis Consortium (DPAC,
\url{http://www.cosmos.esa.int/web/gaia/dpac/consortium}). Funding
for the DPAC has been provided by national institutions, in particular
the institutions participating in the {\it Gaia} Multilateral Agreement.

%\pagebreak
\bibliography{References}
%\pagebreak

\begin{deluxetable}{l c c c c}
\tablecolumns{5}
\tabcolsep0.06in\footnotesize
\tabletypesize{\small}
\tablewidth{0pt}
\tablecaption{ALMA Observations of GQ Lup \label{tab:obs}}
\tablehead{
\colhead{Observation} & 
\colhead{\# of } & 
\colhead{Projected} & 
\colhead{PWV} &
\colhead{Time on} \\
\colhead{Date} & 
\colhead{Antennas} & 
\colhead{Baselines (m)} &
\colhead{(mm)} &
\colhead{Target (min)}
}
\startdata
2015 June 14 & 41 & $16-784$ & 0.6 & 31.3\\
2015 June 15 & 37 & $21-784$ & 0.4 & 31.1  \\
2015 Aug 28 & 40 & $15-1574$ & 1.1 & 35.0
\enddata
\end{deluxetable}

\begin{deluxetable}{l l c c}
\tablecolumns{4}
\tabcolsep0.1in\footnotesize
\tabletypesize{\small}
\tablewidth{0pt}
\tablecaption{$^{12}$CO and $^{13}$CO Model Parameters \label{tab:co}}
\tablehead{
\colhead{Parameter} & 
\colhead{Description} & 
\colhead{$^{12}$CO Best-fit Value} &
\colhead{$^{13}$CO Best-fit Value}
}
\startdata
$M_*$ & Stellar mass ($M_\odot$) & $0.93\pm0.15$ & $1.03\pm0.15$  \\
$i$ & Disk inclination ($\degr$) & $60.3\pm0.4$ & $60.5\pm0.5$ \\
$PA$ & Disk position angle ($\degr$) & $346\pm1$ & $346\pm1$ \\
$r_c$ & Characteristic radius (AU) & $97.6\pm3.7$ & $46.5\pm1.8$ \\
$T_{10}$ & Temperature at 10 AU (K) & $85.5\pm2.5$ & $50.6\pm2.4$ \\
$q$ & Temperature power law index & $0.43\pm0.02$ & $0.38\pm0.04$ \\
log$M_\text{gas}$ & Gas mass (log$M_\odot$) & $-4.72\pm0.03$ & $-3.67\pm0.05$ \\
$\xi$ & Nonthermal broadening line width (km/s) & $0.72\pm0.02$ & $0.55\pm0.02$ \\
$v_\text{sys}$ & Systemic velocity (km/s) & $3.00\pm0.01$ & $3.00\pm0.01$ \\
$\Delta\alpha$ & RA offset ($\arcsec$) & $0.07\pm0.01$ & $0.06\pm0.01$ \\
$\Delta\delta$ & DEC offset ($\arcsec$) & $0.11\pm0.01$ & $0.10\pm0.01$ 
\enddata
\end{deluxetable}

\end{document}